\documentclass[11pt]{article} 
\usepackage{graphicx}
\usepackage{epsfig}

\newcommand {\be} {\begin{equation}}
\newcommand {\ee} {\end{equation}}
\newcommand {\bea} {\begin{eqnarray} }
\newcommand {\eea} {\nonumber \end{eqnarray}}
\newcommand {\eps} {\epsilon}

\newcommand {\ba} {\overline}
\newcommand {\lan} {\langle}
\newcommand {\ran} {\rangle}

\newcommand {\cH}  {{\cal H}}
\newcommand {\cL}  {{\cal L}}
\newcommand {\cN}  {{\cal N}}

\newcommand {\cP}  {{\cal P}}

\newcommand {\bc} {\begin{center}}
\newcommand {\ec} {\end{center}}
\newcommand {\bd}{\begin{displaymath}}
\newcommand {\ed}{\end{displaymath}}

\newcommand {\artanh} {\mbox{artanh}}
\newcommand {\sign} {\mbox{sign}}

\def \form#1 {eq. (\ref{#1}) }
\def \parziale#1#2  {{\partial {#1} \over \partial {#2}}}

\begin{document}
  \author{ Giorgio Parisi \\
	 Dipartimento di Fisica, INFM, SMC and INFN, \\
  Universit\`a di Roma {\em La Sapienza}, P. A. Moro 2, 00185 Rome, Italy. }

\title{Computing the number of metastable states in infinite-range models}
\maketitle
%
\begin{abstract}

In these notes I will review the results that have been obtained in these last years on the computation of the
number of metastable states in infinite-range models of disordered systems. This is a particular case of the problem
of computing the exponentially large number of stationary points of a random function. Quite surprisingly
supersymmetry plays a crucial role in this problem. A careful analysis of the physical implication of
supersymmetry and of supersymmetry breaking will be presented: the most spectacular one is that in the
Sherrington-Kirkpatrick model for spin glasses most 
of the stationary points are saddles, as predicted long time ago.

 \end{abstract}
 \section{Introduction}
 
 The Hamiltonians of  simple (soluble) models of spin glasses are often of the following type \cite{ea}:
 \be
 H=\frac12 \sum_{i,k}J_{i,k}\sigma(i)\sigma(k)\, ,
 \ee
 where the variables $J$ are \emph{quenched} random variables and the spins $\sigma$ take the value $\pm 1$. 
 
 Some of the  mostly studied infinite-range models (that are soluble) are  the Sherrington Kirkpatrick (SK)
 model \cite{sk,MPV,PARISIB,CC} and the Bethe lattice model \cite{MP1Be,FL}.
 \begin{itemize}
	  \item In the SK model the $J$'s are random variables with zero average and variance $1/N$
	  \footnote{They may have either a bimodal distribution or a Gaussian distribution, but these details
	  are irrelevant in the large $N$ limit as soon as the kurtosis is finite}.  In the nutshell the
	  $J$'s form a random Gaussian matrix whose spectrum is given by the semicircle law.
	  \item In the Bethe lattice model we consider a graph with $N$ nodes and $M=2zN$ edges. 
	  A node has in the average $z$ neighbors; in a
	  generic graph, in the limit $N$ going to infinity,
	  the distribution of the number of the neighbors of a node  is a Poisson distribution.  The graph is
	  locally loopless, in the sense that loops of fixed lengths disappear when $N$ goes to infinity:
	   the typical length of a loop is $\log(N)$.  This graph is called a Bethe lattice because
	  the appropriate Bethe approximation is exact on such a lattice.  The Hamiltonian has the same
	  form as before, with the difference that the $J$'s are non-zero only on the edges of the graph. We
	  normalize the  $J's$ in such a way that their variance equal to $z^{-1/2}$. The advantage of this
	  normalization is that in the limit $z \to \infty$ the Bethe lattice model become the SK model.
\end{itemize}

There are many other models and many variations are possible.  For example sometimes the analysis is
simplified by using real-valued variable $\sigma$, that satisfy the spherical constraint
$\sum_{i=1,N}\sigma(i)^2=N$.  There are also systems with three- (or more-)spins interactions.  Other systems
still have two-spins interactions, but they are characterized by a different distribution of the $J$'s, e.g
the $J$'s form a random matrix with a preassigned form of the spectrum; it is also possible to consider the
case where the $J$'s form a non-random matrix \cite{SIN,GRAFFI,potters} as in the sine model
\footnote{The sine model has his own peculiarities and its properties will be not discussed here.}
;
\be
J_{i,k}=N^{1/2}\sin\left({2 \pi i k \over N}\right)\ .
\ee
 
Usually one introduces a $J$-dependent partition function and a $J$-dependent free energy:
\be
\beta f_{J}= \ln(Z_{J}) \ .
\ee
One is interested in computing the probability distribution of $f_{J}$ as function of the $J$'s. However  the free 
energy density 
\be
F_{J}= { f_{J}\over N}
\ee
converges (with probability 1) to a $J$-independent value ($F$) when $N$ goes to infinity.  We must be careful
however in doing the average of the logarithm of $Z_{J}$, because the average of $Z_{J}$
may be much larger of its most likely value.
These models are soluble in the limit of large $N$  in the sense that the free energy $F$ and  other
thermodynamic quantities are computable in this limit.

Spin glasses are also interesting from the point of view of computer science because the computation of the
ground state of a spin glass is an NP complete problem \footnote{This last statement is not so crucial for the
physics: we are usually interested in computing the behavior and the convergence time of an algorithm in the
typical case, while the P-NP classification is based on the worst case.}.  Spin glasses have unexpected
properties (at least in the aforementioned soluble models); the most spectacular is the existence of many
phases at infinitesimal magnetic field, in variance with what happens in standard ferromagnetic models where
there is usually only one phase.  A detailed analytic computation shows that there are an infinite number of
phases or states, more precisely the number of phases increases in an unbounded way with $N$.

I apologize because I am using the word \textit{phase} in an imprecise way \cite{NS}.  I am considering
systems at finite \footnote{Defining the states directly for the
\emph{actual} infinite system is sometime possible, but it is not simple; in other cases it is rather
unnatural: for example in the sine model the Hamiltonian for a system of $N$ spins is not near 
to the Hamiltonian for a systems of $N+1$ spins and we will not discuss further this point.
}, albeit large, $N$ \cite{CINQUE}: roughly speaking a \textit{phase} is a region of phase
space that is separated by other regions by a free energy barrier whose height goes to infinite when the
volume goes to infinity.  Other words, e.g. \emph{structures} or \emph{lumps}, may be more appropriate.
However I will stick here to the more suggestive words \emph{phases} or \emph{states}: the reader should note
that here they are defined in a large \emph{finite} system. Always for finite system one can think that the Gibbs
measure (neglecting a fraction vanishing with $N$) may be partitioned into regions that correspond to the
states. The weight of each state ($w_{\alpha}$) in the decomposition of the free energy should be proportional to the
exponential of the total free energy of the state, i.e. 
\be
w_{\alpha}\propto \exp(-\beta f_{\alpha})\ . 
\ee
The lower lying states   have a weight $w_{\alpha}$ that remains finite when $N$ goes to infinity
\cite{MPV,PARISIB}.
However we are interested in these lecture to the states that have a free energy density higher than the
minimal one (metastable states) and therefore their contribution to the Gibbs measure will be exponentially small. Please notice
that the existence of these states is limited to infinite-range models: real metastable states do not exist for
short range model in finite dimensions \footnote{This is a mathematical theorem that should be taken \textit{cum
granum salis}, as far as the mean life of a metastable state, although finite, may be much larger of the mean
life of the universe.}. The relevance of these finding for short range models is discussed in many places, see
for example \cite{2002}, and it should be done with great care.

In these infinite-range models the internal energy (or the free energy) has have many minima (the
states) and saddle points separating them: we must study the properties of the free energy landscape (that in 
this case is called \textit{corrugated}): we would
like to understand the number and the relative positions of the states.  Moreover, one would like to do this
computation  not only for the
states that have a free energy near to the ground state, but also for those \emph{metastable} states that have
a free energy much higher than the ground state.
 
One can approach this problem, following the suggestion of Monasson \cite{monasson,franzparisi}, in
the framework of purely equilibrium statistical mechanics; we have to compute the partition function
of $m$ replicas, or clones, with some restrictions on the overlaps of the configurations of these
replicas.  The information we need can be extracted (at least in some cases) from the values of this
partition function: this approach is discussed for example in my Les Houches lecture notes of three
years ago \cite{2002} and it will be only briefly summarized here.

Here I will follow a different approach: a state is defined in terms of the local magnetizations and
the local magnetizations satisfy some equations, the so called TAP equations, that become exact (with
probability one) in the limit  $N$ going to infinity \cite{tap,ANDERSON,white}.  Heuristically these equations imply
that there is a long-lived structure: the corresponding local magnetizations satisfy the TAP
equations.  We have to solve the problem of counting the number of solutions of the TAP equations as
function of the free energy and the relative distance of the solutions; eventually we have to
understand how many solutions are minima of the free energy, and how many are $k$-saddles, i.e.
points where the Hessian has $k$ negative eigenvalues,

In the next section of this lecture I will present an heuristic derivation of the TAP equations for
stable and metastable states.  In the rest of the lectures I will compute the average number of
solutions of these equations, which is a very well defined problem from a mathematical point of view.
We will find that the total number of solutions for given values of the $J$ behaves with at large
$N$ as
\be
\exp(N \Sigma_{J})\, .
\ee
Although the most interesting quantity is the \textit{quenched} complexity (defined as the mostly
likely value of $\Sigma_{J}$) here for simplicity we will limit ourself to the computation of the so
called \emph{annealed} complexity:
\be
\exp(N \Sigma))= \ba{\exp(N \Sigma_{J})}\ ,
\ee
where the average is done over the different choices of the $J$'s variables.  Depending on the model
and on the parameters of the models the quenched complexity may be strictly smaller or equal to the
annealed complexity in the infinite-$N$ limit.

Computing the complexity is a particular case of the more general problem of finding all the stationary points
of a random function of $N$ variables, that in this case is the TAP free energy.  In section 2 I will
present the derivation of the TAP equations for the magnetizations of a metastable state and I will discuss
their physical meaning.  In section 3 I will present some simple considerations on the number of solutions of
the TAP equations.  In section 4 I will show the general strategy for computing the number of solutions of the
TAP equations as function of the corresponding free energy.  In section 5 I will show that this problem is
invariant under the action of a Fermionic symmetry, a supersymmetry in short; this unexpected and astonishing
results was derived a few years ago in \cite{juanpe} and it was the key of all the progresses of there recent
years.  In section 6 I will discuss the possibility of supersymmetry breaking and its implications, the most
spectacular one is that most of the stationary points are one dimensional saddles, as predicted long time ago
\cite{tap,ANDERSON}.  In section 7 I will present the explicit computation of the complexity in the SK model,
sketching the different steps that have to be done.  In section 8 I will just mention some of the problems
that are present in computing the quenched complexity.  Finally in section 9 I will summarize the results and
mention some open problems.

I believe that most of the computations that I present can be done
in a rigorous mathematical way using existing techniques in a careful way (the computation of the quenched
complexity may be much more difficult).  I hope that these lectures will be
an encouragement to do such a rigorous computation.

\section {The TAP Equations}

I will derive the TAP equations on a random Bethe lattice \cite{tap,MP1Be,FL,white} where the average number
of neighbors is equal to $z$ and each point has a Poisson distribution of neighbors~\footnote {One can
consider also a generalized random Bethe lattice where each site has a fixed coordination number: the
treatment is similar, but slightly different.}: the SK limit will be done later.

Let us consider a node $i$; we denote by $\partial i$ the set
of nodes that are connected to the point $i$. 
With this notation the Hamiltonian can be written
\be
H=\frac12 \sum_{i}\sum_{k\in \partial i}J_{i,k}\sigma(i)\sigma(k)\, .
\ee

We suppose that for a given value of $N$ the system is in a state labeled by $\alpha$ and we suppose such a
state exists also for the system of $N-1$ spins when the spin $i$ is removed.  Let us call $m(i)_{\alpha}$ the
magnetization of the spin $k$ and $m(k;i)_{\alpha}$ the magnetization of the spin $k$ when the site $i$ is
removed.  Two generic spins are, with probability one, far on the lattice: they are at an average distance of
order $\ln(N)$; it is reasonable to assume that in a given state the correlations of two generic spins are
small (a crucial characteristic of a state is the absence of infinite-range correlations).  Therefore the
probability distribution of two generic spins is factorized (apart from corrections vanishing in probability
when $N$ goes to infinity) and it can be written in terms of their magnetizations.

After a simple computation we find for a given state labeled by $\alpha$ (the label $\alpha$ will not be
indicated to lighten the notation):
\be
m(i)=\tanh\left( \sum_{k\in \partial i} \artanh(\tanh (\beta J_{i,k})m(k;i))\right) \label{MAG}\ .
\ee
In a similar way we finds that the cavity magnetizations $m(k;i)$ are related to the magnetizations $m(k)$
and $m(i)$ by the relations:
\be
m(k)=\tanh(\artanh(m(k;i)) + \artanh(\tanh (\beta J_{i,k})m(i)))  .
\ee
In the case of  the Bethe lattice the usual strategy is to write the equation directly for the cavity
magnetizations. We obtain (for $l\in\partial i$):
\be
m(i;l)=\tanh \left( \sum_{k\in \partial i; k\ne l} \artanh(\tanh (\beta J_{i,k})m(k;i)) \right) \ .
\ee 
Following this strategy we remain with $Nz$ equations for the $Nz$ cavity magnetizations $m(i;k)$;
 the true magnetizations ($m(i)$) can be computed at the end using equation (\ref{MAG}).

In the SK limit (i.e. $z\approx N $), it is convenient  to take  advantage of the fact that the 
$J$'s are proportional to
$N^{-1/2}$ and therefore the previous formulae may be partially linearized. If we throw away   irrelevant
terms, the previous equations become (as soon as $z$ is large):
\bea
m(i)=\tanh\left( \sum_{k\in \partial i} \beta J_{i,k}m(k;i)\right) \ , \label{PRETAP}\\
m(k)=m(k;i)+\beta (1-m(k)^2) J_{i,k}m(i) \ .
\eea
If we eliminate all the  terms that vanish when $N \to\infty$, 
these two equations can be simply rewritten  as
\bea
m(i)=\tanh(\beta h(i)), \\  h(i)\equiv \sum_{k\in \partial i}  J_{i,k}m(k;i)=
\sum_{k\in \partial i}  J_{i,k}m(k) - \beta m(i)\sum _{k\in \partial i}J_{i,k}^2(1-m(k)^2) \ .
\eea
In the SK limit the points in $\partial i$ are all the points different from $i$.
Using the law of  large numbers
we can write
\be
m(i)=\tanh(\beta h(i)), \ \ \ \ \ h(i)\equiv \sum_{k\ne i}  J_{i,k}m(k) - \beta m(i)(1-q) \label{TAP}\, ,
\ee
where $q$ is defined as
\be
q={\sum_{i} m(i)^2\over N}\ .
\ee

These equations are different from the naive mean field equations
\be
m(i)=\tanh(\beta h(i)), \ \ \ \ \ h(i)\equiv \sum_{k\ne i}  J_{i,k}m(k)\ .
\ee
The naive mean field equations can be derived using a variational approach where the probability is
supposed to be factorized.  In this way one neglects the weak correlations among the $\sigma$'s and this
leads to a wrong value of the free energy at non-zero temperature~\footnote{The naive mean field
equations are exact equations for another model that is different from the SK model.  Unfortunately
they have not been so well studied in recent times.}.  On the Bethe lattice the argument is quite clear.
The spins that belong to $\partial i$ \emph{are} \emph{correlated} because they are at distance $2$ on the
lattice; on the contrary the same spins \emph{are not} \emph{correlated} when the spin at $i$ is removed 
(they are at distance of $O(\ln(N))$ and 
therefore the factorization approximation can be done for the spins in $\partial i$ only when the
spin at $i$ is removed.

The same argument can be formulated in a slightly more different way, that can be  
generalized to more complex cases.  It is the cavity method, where one connects the magnetizations
for a system of $N$ spins with the magnetizations for a system with $N-1$ spins.  In absence of the
spin $i$, the spins $\sigma(k)$ ($k \in \partial i$) are independent from the couplings $J_{i,k}$.
If we linearize the equations for the Bethe lattice, we obtain:
\bea
 m(i)=\tanh\left( \sum_{k\in \partial i}J_{i,k}m(k;i)\right)\, \\
 m(k;i)=\tanh\left( \sum_{l\ne i}J_{k,l}m(l;k,i)\right)\ .
\eea
 Here $m(l;k,i)$ is the magnetization of the spin at $l$ in absence of the spins at $k$ and at $i$.
 A detailed computation shows that in this case, neglecting terms that go to zero (in probability) when $N$
 goes to infinity, we can substitute $m(l;k,i)$ with $m(l;k)$. Indeed the nodes $i$ and $l$ are at distance two
 when the node at $k$ is present: however, when $k$ is removed, they are far away in the lattice.
 In this way we recover eq. (\ref{MAG}).
 
Now we will concentrate on the SK model and we will use the formulation where we have  $N$ variables and $N$
equations, i.e. the original TAP equations, eq. (\ref{TAP}).
 The  corresponding  total free energy is 
 \be
 f[m]=-\frac12 \sum_{i,k}m(i)J_{i,k}m(k)-\frac{\beta}{4}\sum_{i}(1-m(i)^2)- T \sum_{i}S(m(i)) , \label{FREE}
 \ee
 where 
 \be
 S(m)=-\frac{1+m}{2}\ln\left(\frac{1+m}{2}\right)-\frac{1-m}{2}\ln\left(\frac{1-m}{2}\right)
 \label{ENTROPY}\ .
 \ee
 The free energy is stationary: in other words the TAP equations can be written as
 \be
 {\partial f[m]\over\partial m(i)}=0 \ .
 \ee
 However, as we shall see, that there is no variational  principle for the free energy: the equilibrium value of 
 the free energy is \emph{not} the absolute minimum of the free energy.
 The Hessian of the free energy can be written as
 \bea
 \cH_{i,k}\equiv{\partial^2f\over \partial m(i)\partial m(k)}=\\
 -
 J_{i,k}+\delta_{i,k}\left(A(i)^{-1}+\beta(1-q)\right)+\frac{2\beta}{N}m(i)m(k)\ 
 , \label{HESSIAN}
 \eea
 where we have  used the notation
\be
A(i)\equiv{\beta (1-m(i)^2)}\ .
\ee
 The relation
 \be
\beta (1-q)={\sum_{i=1,N}A(i)\over N}\ .
\ee 
may have not escaped  the attention of the experienced reader.  It is rather important because it can be
used to prove that in the large-$N$ limit the Hessian does not have a non-zero fraction of negative
eigenvalues \cite{PLE}.
 
As we shall see later, we must resist  the strong temptation of throwing away the last term that is a factor
$1/N$ smaller than the others.  Indeed, once that we have written the free energy, the Hessian is exactly given by the
previous formula. Before throwing away a term, you must ponder all the consequences of your act: you may
discover 24 years later that such an action was not justified \cite{bm,R2}!

We must stress that there is an important consistency condition: the correlation among different
spins must vanish in the large $N$ limit.  Fortunately they can be computed in this framework.
After some computations we find that the correlations are small enough and the approach may be consistent only
if the stability condition
 \be
 R\equiv{\sum_{l=1,N}A(i)^2\over N}\equiv \beta^2 {\sum_{l=1,N}(1-m(i)^2)^2
 \over N}\le 1\label{STAB}
 \ee
 is satisfied. Indeed we find that
 \be
 \chi_{q}\equiv{\sum_{i,k}\lan\sigma(i) \sigma(k)\ran_c^2 \over N}={1 \over 1-R}\ . 
 \ee
If $R>1$ the whole approach does not make sense.

Using the spectral properties of the matrix $J$ (for $N$ going to infinity the spectral density of
this random matrix is a semicircle with boundaries at $\pm 2$) we find that the paramagnetic
solution $m(i)=0$ is always a minimum of $F$.  However the paramagnetic solution is no more in the
stable region as soon as $\beta>1$, i.e. in the whole low-temperature region.  For
$\beta>1$ the paramagnetic solution $m(i)=0$ should be disregarded because is not stable.

The physical picture is clear.  At low temperature the free energy has a deep minimum at the origin, however
the physical relevant region does not contain the origin and the physical relevant minimum is the lowest
minimum inside the stable region ($R\le 1$).  We should also notice that the equivalence among this approach
(with $N$ equations) and the cavity approach (with $O(N^2)$ equations) holds only if the stability condition
is strictly satisfied and it is incorrect when the stability condition is not satisfied.

In the more familiar case of a ferromagnet, the paramagnetic solution is no more a global minimum in the 
low-temperature case, while in spin glasses the paramagnetic phase is always  the  formal global minimum of the free energy.
This difference has a profound influence on the behavior of the specific heat  (in the mean field
approximation): 
\begin{itemize}
\item In the ferromagnetic case the specific heat in the low-temperature phase is \emph{larger} than the
analytic continuation of the specific heat computed in the high-temperature phase. This corresponds to the
fact that in the low-temperature phase the system explores new regions of phase space; this is the natural
behavior when we have a variational approach without constraints.

\item In spin glasses the situation is the opposite: the specific heat in the low-temperature phase is
\emph{smaller} than the analytic continuation of the specific heat computed in the high-temperature phase.
This corresponds to the fact that the system is frozen in the low-temperature phase and here the paramagnetic
solution is only an illusion.  This behavior is the effect of the constraints in the standard variational
approach.
This strange behavior is natural in the variational replica approach where the free energy is maximized (for
good reasons \cite{Guerra}) and not minimized.  It is interesting to note that experimentally in  structural  glasses  the
specific heat is discontinuous and it decreases when we cross the critical temperature coming from the high-temperature phase.
\end{itemize}
\section{A simple analysis of the solutions of the Bethe equations} 

In the simplest approach we assume that in the low-temperature phase there is just one solution to the
equations (\ref{TAP}) (neglecting the paramagnetic one).  The field $h(i)$ can be written using the cavity
equation as the sum of $N-1$  uncorrelated random terms with zero average \footnote{In doing these evaluations
it is more convenient to use the equations \ref{PRETAP}; we could also start from the equations \ref{TAP}; the
results would be the same and the derivation would be slightly more involved.}: the $J_{i,k}$ are uncorrelated
with the $m(i;k)$ so that $J_{i,k}m(i;k)$ has  zero average and the contributions with different values of
$k$ are uncorrelated.  It is evident that the average value
of $m(i;k)^2$ must be equal to $q$.  The central limit theorem tell us that, when we change the $J_{i,k}$,
$h(i)$ is a random variable with
variance $ q$ \cite{MPV}.

We will now use the relation:
\be
\ba{m(i)^2}= q\, ,
\ee
where here the bar denotes the average over the couplings $J_{i,k}$. In this way
we arrive at the so called replica symmetric equations for $q$:
\be
q=\int d\mu_{ q}(h) \tanh(\beta h)^2 \label{RS}\, ,
\ee
where $d\mu_{ q}(h)$ denotes the  Gaussian probability distribution with variance $q$.

A simple analysis shows that for $T<1$ there is a non-trivial solution with $q\ne 0$ (the solution with $q=0$ is the
paramagnetic solution that is unstable for  $T<1$) . However, if we compute the
entropy at low temperature, we find the astonishing result that it is negative \cite{sk}. This is possible because the total
entropy  is \emph{not} the sum of the single-spin entropy and its explicit expression does 
 \emph{not} have an evident sign. Indeed, using the TAP free energy, we find that:
 \be
 S= \sum_{i}S(m(i)) - {\beta^2 \over 4} (1-q)^2 \ .
 \ee
 In the limit $T$ going to zero, if $q \to 1$ the first term vanishes, but the second is strictly negative, if 
 \be
 \lim_{T \to 0} \beta (1-q) >0
 \ee
 and this is what happens as a simple consequence of  equation (\ref{RS}).

Something goes wrong.  The solution of the puzzle was found by De Almeida and Thouless \cite{DAT}
who discovered that as soon as $T<1$ also the $q\ne 0$ solution is unstable, i.e. it does not
satisfy the stability condition (\ref{STAB}) and no stable solutions are available.  After some
hesitations, at the end of the day consensus was reached on the fact that the hypothesis of having only
one non-trivial solution was inconsistent: we must take into consideration  the existence of an infinite number
of solutions of the TAP equations: this situation is called replica symmetry breaking
\cite{MPV,PARISIB,CC}.

The central quantity in the theory becomes now the set of all the solutions of the TAP equations and
their corresponding free energies \cite{bm,dd1,tana,ddy,potters2}.  In the simplest case we assume
that the solutions are
uncorrelated among themselves and that they are uncorrelated with the free energy.  In other words
we assume an approximate factorization property for the joint probability distribution of the total
free energy $f_{\alpha}$ and the corresponding magnetizations $m^\alpha(i)$, where the index
$\alpha$ labels the different solutions of the TAP equations \cite{MP1Be,2002}.  More precisely we
suppose that
\be
P_{i}(f+\tilde f,m)=\cP(f+\tilde f)P_{i}(f,m)\ ,
\ee
where $\tilde f$ is  a quantity of order 1 and 
\be
\int df \cP(f)= \infty \ ,
\ee
because this integral  is the total number of solutions of the TAP equation.

We now assume the probability $\cP(f)$ is approximatively given by
 \be
 \exp(N\Sigma(F))
 \ee
 where $F=f/N$. This is a natural extension of our previous hypothesis on the complexity. The total complexity
 is given by
 \be
 \Sigma=\max_{F}\Sigma(F) \ .
 \ee
 In a similar way we assume that 
 \be
 P_{i}(f,m)=\cP_{i}(F,m)\ .
 \ee
 
Under these assumptions the probability  of  finding a solution around $f=F\, N$ can be written as  
 \be
 \cP(f+\tilde f)=\cP(f)\exp( y(\tilde f -f)) \propto \exp(y(\tilde f -f)) , \ \ \  
 \ee 
 where $\tilde f$ is a  still quantity of order 1 and 
 \be
 y={d\Sigma\over dF} \ .
 \ee
In this framework   is natural to assume that, when the free energy is near to the minimal free energy (that we
will denote by $f_{0}$),
the probability of finding a solution with  total free energy given by $f_{0}+ \tilde f$   behaves like 
\be
\cP(\tilde f) \propto\exp(y_{0} \tilde f) \ .
\ee
Also here $\tilde f$ is a  still quantity of order 1 while $f_{0}$ is a quantity of order $N$.
I will use the notation  $y=\beta m$ and $y_{0}=\beta m_{0}$, where quantities $m$ and  $m_{0}$ are
{\sl not}  magnetizations, (I will stick to
this traditional notation, hoping that there will be no confusion).

In order to make further progress and perform  explicit computations, we can write equations that relate the distribution of
the magnetizations for a system with $N$ points to the distribution of
the magnetizations for a system with  $N+1$ points. We have to make the crucial
hypothesis that there is a one to one correspondence between each solution of  the TAP equations for  $N$
spins and for 
$N+1$ spins; as we shall see later this hypothesis fails at high value of the free energy, but it is reasonable for values of the free
energy not too far from the minimal one.

Let us consider  the solutions that have a free energy density $F$. We need to know only the
value of $m$. After some computations, described  in my earlier Les Houches notes \cite {2002} one finds that
\be
dP_{N+1}(h)\propto  d\mu_{ q}(h) \tanh(\beta h)^2 \exp(\beta m \Delta f(h))\, ,
\ee
where 
\be
m_{N+1}=\tanh(\beta h_{N+1})\, ,
 \ee
 and
 \be
 \Delta f(h) =\log(\cosh(\beta h)) \ .
 \ee
 Consistency requires that 
 \be
 q=\int dP_{N+1}(h) (\tanh(\beta h_{N+1}))^2 \label{ONEQ} \ .
 \ee
 
 In this way we determine $q$ but $m$ is a free parameter. How can one determine $m$?  It happens that it
 is possible to compute  the free energy density $F$ that corresponds to a value of $m$. This is done in the
 following way.
  
We can define a thermodynamic potential  $\Phi(m)$ such that
 \bea
 {\partial \Phi(m) \over \partial\beta  m} =\Sigma(F(m))\, \\
 {\partial(\beta m\Phi(m))\over \partial \beta m}=F(m)
 \eea
 In other words $\Phi(m)$ is the Legendre transform of the complexity:
 \be
 \beta m\Phi(m)=\beta m F(m)- \Sigma(m) \ .
 \ee
 You may have a feeling of \emph{d\'{e}j\`{a} vu}: indeed we recover a more familiar setting
 \cite{VEVH} if we would do the
 following substitutions:
 \bea
 F \to \mbox {internal energy} \, \\
 \beta m \to \beta \, \nonumber \\
 \Phi \to \mbox{free energy}\,\nonumber \\
 \Sigma \to \mbox{entropy.}\nonumber
 \eea
 It should be now clear why in some circles the complexity is called  the configurational entropy 
 \cite{CE}.
 
If we know the the potential $\Phi(m)$, we know all we need to reconstruct everything, in
particular the dependence of the complexity on $m$ or on $F$.  Fortunately an analysis, that it would be
too long to describe here \cite{2002}, shows that $\Phi(m)$ has a simple explicit
expression \footnote{The same expression for the free energy can be obtained using the traditional replica
approach.}:
 \be
 \beta^2 (1-q^2)+m^{-1}\int_{-\infty}^{+\infty} d\mu_{q}(h) \cosh(\beta h)^m \log(\cosh(\beta h))\ .
 \ee
 A similar, but more complex expression for $\Phi(m)$ can also be obtained on the Bethe lattice \cite{MP1Be}.
 
 It turns out that the free energy is stationary with respect to $q$; indeed the correct value of $q$ is
 obtained by \emph{maximizing} the free energy with respect to $q$: the previous equation (\ref{ONEQ})
 corresponds to imposing the condition
 \be
 {\partial \Phi \over \partial q}=0
 \ee
 and the sign of the second derivative is such that the potential $\Phi$ is a maximum with respect to $q$.
 
We now face the problem of computing the minimal free energy (sometimes also called the ground state free energy. 
This is now an easy job if we make the
natural hypothesis that the complexity vanishes when the free energy density becomes equal to the ground
state.  In other words the value of $m_{0}$ corresponding to the ground state energy is given by the condition
  \be
  \Sigma(m_{0})=0\ .
  \ee
  However the previous equations tell us that  
  \be
   \Sigma(m)={\partial\Phi(m) \over\partial m} \ .
    \ee
   
In other words the condition of zero complexity at the ground state, implies that the potential
$\Phi(m)$ is stationary with respect to $m$.  More precisely in  order to compute the free energy of the
ground state we have to maximize the potential $\Phi$ with respect to both $q$ and $m$: for the ground state
we have an \emph{inverted} variational principle.

The formula for the potential $\Phi(m,q)$ is the one that has been obtained directly with the
algebraic replica approach long time ago.  Using Guerra's variational approach \cite{GUERRA} it can be
shown from general principles that the maximization of the potential $\Phi(m,q)$ gives the correct
result \cite{TALA}.  It is remarkable that the same formulae, that have been originally derived for
the ground state, can be  used for the computation of the complexity.  This is one of the most
astonishing result of Monasson's approach that we have just described.

However in the previous derivation we have made the crucial hypothesis of the existence of a  one to one correspondence of the
solutions of the TAP equations with $N$ and $N+1$ spins.  At first sight this hypothesis is witless, because we
know that the total number of solutions increase with a factor $\exp(\Sigma)$ when we go from $N$ to $N+1$.
As we shall see later this apparently witless hypothesis may remain valid for some values  of $F$.
 
The rest of this lectures will be devoted to discussing when and why such an hypothesis is correct and to
present alternative approaches in the case where it fails.  The relations among the different approaches are
non-trivial and they come out from some very interesting mathematics.
 
 \section{The direct approach: general considerations}
 
 Before starting  the direct computation of the number of solutions of the TAP equations, I will present some
 general considerations on  this kind of problems and I will write down some formulae that   are valid
 in a more general case \cite{bm,dd1,tana,ddy,potters2}.

I will consider a function $f[x]$ where here $x$ (among square brackets) denotes the set of $N$ variables $x(i)$,
with $i=1,N$. If the function $f$ factorizes in the product of $N$ non-linear functions, each dependent on only
one variable $x(i)$, the analysis is rather simple; in the case where each function has 3 stationary
points, the total number of stationary points of the function $f[x]$ is $3^N$.

In the case where all the variables interact and no factorization is possible, the situation is rather
complex; also in this case it is quite possible that we have an exponential large number of stationary points (i.e. points where
${\partial f / \partial x}=0$) .  A very interesting question is the computation of the properties of these
points, e.g their number and there relative location in space.  If the number of the points diverges
exponentially with $N$, methods coming from statistical mechanics may be very useful.
 
As we have seen in previous sections this problem is very relevant for statistical mechanics.  Indeed if the
function $f$ is the free energy or the internal energy of a system, its local minima correspond to stable or
metastable states at non-zero (or zero) temperature. The relative height and position of other stationary
points, e.g. saddles, control the time from escaping from the metastable states and the approach to
equilibrium.
 
When the function $f$ is the free energy, the shape of this function is called the free energy landscape.  If
the number of stationary points is very high, e.g. increases exponentially with $N$, we have already seen  that the landscape is
called corrugated. A system with a corrugated landscape is a glassy system: indeed  these investigations are often done
to study glassy systems like spin glasses or structural glasses, although they could also be relevant for
string theories.
 
The first quantity that we have to compute is the total number of stationary points
\be
\cN_{T}\equiv \sum_{\alpha\in C} 1
\ee
or the total number of minima
\be
\cN_{m}\equiv \sum_{\alpha\in C_{m}} 1 \, ,
\ee
where the $C$ denotes the set of stationary points and  $C_{m}$ those stationary points that are actually minima.

Very often one is interested also in the distribution of the values of the function $f$ at the stationary points (indeed 
there may be a physical interest in knowing 
 the stationary points with the lowest value of $f$). It is therefore
convenient to consider the two functions:
\bea
\cN(F)=\sum_{\alpha\in C} \delta(F-f[x_{\alpha}])\\
Z(w)=\sum_{\alpha\in C}\exp (-w  f[x_{\alpha}]) =\int dF \cN(F) \exp(-w N F) \ .
\eea
For large $N$ we expect that
 \be
 \cN(F) \approx \exp (N  \Sigma(F))\ ,
 Z(w) \approx \exp (-N \Phi(w))\ ,
 \ee 
 where $ \Phi(w)$ and $ \Sigma(F)$ are related by an Legendre transform (exactly as the free energy and the
 entropy in the usual formalism):
 \bea
 {\partial \Sigma(F) \over \partial F}= w(F)\ ,\\
 \Phi(w(F))= \Sigma(F)-w(F)F \ .
\eea
If we are able to compute the function $Z(w)$ we can reconstruct, by an inverse
Mellin transform, the function $\cN(F)$ that codes the distribution of the values of the stationary points. In
particular we have that
\be
Z(w)|_{w=0}=\exp(N\Sigma)
\ee
Everything is very similar to what we have done in the previous section, with the difference that here we use 
a \emph{synchronic} approach (we look to what happens at a given value of $N$), while in the previous section we were
using a \emph{diachronic} approach (we were following the solutions of the TAP equations when $N$ was changing).
Here we do not need to make an explicit hypothesis on the continuity of the set of solutions with respect to
$N$; however we shall see later how this continuity hypothesis can be reformulated in the synchronic approach.

For the time being we will concentrate our attention on the quantity $Z(w)$ that looks like a partition function,
defined on an unusual space, i.e. the set of all the stationary points.  It is important to remember that the
function $f[x]$ that appears in the weight is the same function that is used to determine the points of the
sum, and this will be the origin of many identities.

In order to make further progress let us assume ( happens for a generic function and for a generic
stationary point ) that the determinant $D[x]$ of the Hessian matrix $\cH$ of the function $f[x]$ is non-zero at the
stationary points, i.e.
 \be
 D[x]\equiv \det |H| \ne 0 ,\ \ \ \  \cH_{i,k}={\partial^2f\over\partial x(i)\partial x(k)} \,
 \ee
 when ${\partial f / \partial x}=0$.  In one dimension this condition corresponds to assuming that all stationary points
 are minima or maxima (no flexes). If this happens we can define the index ($I[x]$) of a stationary
 point; it is the number of negative eigenvalues of the Hessian; a minimum has index 0 while a maximum has an
 index $N$.  An useful and evident relation is:
 \be
 (-1)^{I[x]}=\sign(D[x])=D[x]/|D[x]|
 \ee
 In the following we will consider for simplicity functions that are defined on a topologically trivial domain (e.g. an
 hypercube, the interior of a sphere or the whole space $R^N$)  such that the function $f[x]$ diverges to plus 
 infinity  at the boundary.
 
 Under these conditions it may be convenient to compute also the function
 \be
 \tilde\cN(w)=\sum_{\alpha\in C}(-1)^{I[x_{\alpha}]} \exp(-w f[x_{\alpha}]) \ ,
 \ee 

 Skipping technicalities, the celebrated Morse theorem states that 
 \be
 \tilde\cN(w)|_{w=0}=\sum_{\alpha\in C}(-1)^{I[x_{\alpha}]}=1\ .
 \ee
 
 In the  one-dimensional case the Morse theorem states that, if $\lim _{x\to \pm \infty}f[x] = +\infty$, the
 number of minima minus the number of maxima is exactly equal to one, as  follows from elementary
 considerations.
 
 The previous equations tell us that the behaviors of $\tilde\cN(w)$ and of $\cN(w)$ are quite different,
 however it may be convenient to start with the computation of $\tilde\cN(w)$ that is much simpler than the
 other.  Later on we will show how the computation of $\cN(w)$ can be recovered.  Indeed a direct estimate can
 be done \cite{XXX}, following some ideas of Kurchan \cite{ps3,jorge1}, using d'Oresme's identity 
 \be
 |A|=(A^2)^{1/2} \ .
 \ee
 
 In order to compute $\tilde\cN(w)$  we firstly notice that the sum over all the stationary points can be
 written as
 \be
 \sum_{\alpha\in C}=\int dx |D[x]| \delta \left({\partial f\over\partial x}\right)
 \ee
 Therefore 
 \bea
 \tilde\cN(w) = \int dx D[x]\prod_{i=1,N}\delta(f_{i}[x])  \exp(-wf[x]) \, \\
 \cN(w) = \int dx |D[x]|\prod_{i=1,N}\delta(f_{i}[x])  \exp(-wf[x])\, 
 \eea
 where from now on we shall use the notation
 \be
 f_{i}[x]={\partial f[x]\over \partial x(i)} \ \ \ \ \ \ f_{i,k}[x]={\partial^2 f[x]\over \partial x(i)\partial
 x(k)}.
 \ee
Now the computation of $N(w)$ is technically difficult because of the presence of the modulus of the
determinant. It should be clear now why the  function that can be computed more easily is $\tilde N(w)$.
 
In the rest of these notes we will stick to the computation of $\tilde \cN(w)$.
Now, in order to do further manipulations,  it is useful to write the determinant as an integral over 
Fermionic (anticommuting) variables using the formula.
\be
\int \prod_{i}(d\bar \psi(i)d \psi(i)) \exp(\sum_{i,k}A_{i,k}\bar \psi(i)\psi(k))= \det |A| \ .
\ee
If we also use the integral representation for the delta function we
obtain that 
\bea
\tilde\cN(w)=\int d\mu[X] \exp(-S[X]) \, \\
S[X]=\sum_{i}\lambda(i)f(i)[x]+\sum_{i,k}\bar\psi(i)\psi(k)f_{i,k}[x]+w f[x] \nonumber \ ,\\
d\mu[X] \equiv \prod_{i} dx(i)d\psi(i)d\bar\psi_{i }d\lambda(i) \, \label{BOH}
 \eea
 where $X(i)$ denotes a point of the superspace with coordinates  $x(i)$'s, $\bar \psi(i)$ $\psi(i)$ and
 $\lambda(i)$; the integral over the variables $\lambda$ is done in the complex plane from $-i\infty$ to
 $+i\infty$.
 
 \section{The supersymmetric formulation}
 The computation of $\tilde\cN(w)$ may be simplified if we notice that both the action $S$ and the measure $d\mu$
 are left invariant under the following supersymmetry \footnote{I use the word supersymmetry in a somewhat
 improper way; the symmetry we have introduced  contains  an anticommuting parameter, and it a {\it Fermionic} symmetry, but in
 the standard terminology a
 {\it bona fide} supersymmetry group should contain also the generator of the space translations, that is not 
 the case here.} that was introduced in \cite{juanpe};
 \be
 \delta x(i)=\eps(i) \psi(i)\, \ \ \delta \lambda(i)=- \eps(i) \; {w}  \psi(i)\ ,\ \ \delta \bar{\psi}(i)=\eps(i)
 \lambda(i)\,  \ \ \ \delta {\psi}(i)=0\ .
 \ee
 The appearance at $w\ne 0$ of a supersymmetry, of the BRST type \cite{brs,tito}, is unexpected, however it is
 well known that the Morse theorem, which states that $\tilde\cN(w)_{w=0}=1$, can be proved in a neat way using the
 supersymmetric formalism.
 
As usual supersymmetry \footnote{This supersymmetry does not have a very simple geometrical
interpretation in superspace also because the transformation depends on $w$; it is not clear to me
if the whole approach may be reformulated in such a way that the supersymmetry has a more
geometrical meaning.} implies identities among correlations of different quantities; in this
particular case one finds after some computations \cite{juanpe,CGMP}
 \bea
 \lan \bar \psi(i)\psi(k)\ran =\lan x(i)\lambda(k)\ran \ ,\\
 w \lan \bar \psi(i)\psi(k)\ran = \lan \lambda(i)\lambda(k)\ran \ .
 \eea
 
 The derivation of the Fermionic symmetry is purely formal, therefore it would be nice to derive directly the
 Ward identities from the original equations without using too much the functional integral representation.  We
 need to assume that when we make a small change in the equations, there is a one to one 
 correspondence between the solutions before and after the perturbation. We also need to assume, but this
 assumption is related to the previous one, i.e. that the quantities that appear in the Ward identities are
 well-defined.
 
 In order to understand better the Ward identities \cite{R3} we  notice that
 \be
 \lan \bar \psi(i)\psi(k)\ran =\lan(\cH^{-1})_{i,k}\ran \, ,
 \ee
 where $\cH$ is the Hessian matrix. The computation of the correlations of the $\lambda$'s, that, in the
 nutshell,  are 
 Lagrange multipliers, can be obtained by introducing the generating function:
 \bea
 Z[h]=\int d\mu(X) \exp\left(-S(X)+\sum_{i}h(i)\lambda(i)\right)= \\
 \int dx   D[x]\prod_{i=1,N}\delta\left(f_{i}[x]-h(i)\right)  \exp(-wf[x]) \ . \label{PARTMOD}
 \eea
In other words we count the number of solutions of the equations 
\be
 f_{i}[x]=h(i) \ ,\label{EQMOD}
 \ee
weighted with $\exp(-wf[x])$, (not with $\exp(-w(f[x] -h\cdot x)$). As soon as $h$ is different from zero, the 
constraints on the variables $x$ and the weight are no more related and supersymmetry does no more hold.

 Let us make the crucial hypothesis that for infinitesimal $h$ there is a one to one correspondence between the
 solutions of the equations
 \be
 f_{i}[x]=h(i)
 \ee
 at zero $h$ and at non zero $h$ and that the solution at non zero $h$ can be computed by a well-defined perturbative
 expansion in $h$ around $h=0$ \cite{CLPR,CGP5}.
 It is now evident that at $h=0$
 \be
 \lan x(i)\lambda(k)\ran=\lan {\partial x(i)\over \partial h(k)}\ran =\lan(\cH^{-1})_{i,k}\ran=\lan \bar
 \psi(i)\psi(k)\ran \ .
 \ee
 And in this way we recover the first Ward identity, the one where $w$ does not appear.
 
 In a similar way if we work up to the second order in $h$ in the free energy, we can derive the first order result 
 for the stationary point
 \be
 \delta x(i)\equiv x(i)[h]-x(i)=\sum_{k}(\cH^{-1})_{i,k} \delta h(k) +O(h^2) \ .
 \ee
 The variation of the function $f$ starts at the second order in the perturbation
 \be
 \delta f=\frac12\sum_{i,k}\delta x(i)\cH_{i,k} \delta x(k)=
 \frac12\sum_{i,k}\delta h(i)(\cH^{-1})_{i,k} \delta h(k).
 \ee
 Putting everything together we get:
 \be
 \lan \lambda(i)\lambda(k)\ran={\partial^2 Z[h]\over \partial h(i)\partial h_k}|_{h=0}=
 w \lan(\cH^{-1})_{i,k}\ran=\lan \bar \psi(i)\psi(k)\ran \ .
 \ee
 which is the second Ward  identity, i.e. the one that depends on the $w$.
 
 Supersymmetry is an efficient way include relations of purely geometrical
 origin between 
 the correlations of the  Bosonic variables, which play the role of Lagrangian multipliers, and 
 the correlations of the
 Fermionic variables, which have been used to evaluate the inverse of the Hessian matrix and its determinant.
 
 \section{Spontaneous supersymmetry breaking}
 
 It may seem strange that I am speaking of spontaneous breaking of the supersymmetry just after having proved
 that supersymmetry encodes simple identities among various quantities
 \cite{R3}.  However these identities were derived
 under the  critical hypothesis of the existence of a one to one correspondence of the solutions of the
 equations after and before a perturbation.  This is reasonable only if the generic solution of the system is
 not near to a bifurcation and this is possible if the Hessian matrix does not have a zero or a quasi-zero
 mode.  Everything was based on the hypothesis that the Hessian matrix at the stationary points had no zero-mode.

 If we
 average the function $f$ inside a
 given class, we may integrate over  regions where zero-modes are always present and our assumptions 
 fail. Moreover in such a case $\cH^{-1}$ is infinite and we have problems defining the expectation
 value of $(\cH^{-1})_{i,k}$. 
 
 Still worse, if we write 
 \be
 \lan O \ran =\sum_{\alpha}w_{\alpha}(h) \lan O_{\alpha}(h)\ran
 \ee
 when we take the derivative with respect to $h$ of this quantity, we have to take into account in an explicit
 way the fact that the set of solutions labeled with $\alpha$ changes may with $h$.  This looks like a
 \emph{cauchemar} but it is possible to do explicit computations in the framework of the cavity method of the
 previous sections \cite{CGP5}.  One finds that the results are different from the usual supersymmetric
 approach and they coincide with the one obtained by assuming that supersymmetry is spontaneously broken
 \cite{CGP5}.

 We could try to avoid the problem by adding a small term in the weight in order to suppress the contribution
 coming from stationary points with zero modes, but this extra term breaks explicitly the supersymmetry and it
 is not surprising that in the thermodynamic limit supersymmetry breaking may survive when we remove the extra
 term \cite{RI0}.  However computation can be done in presence of this extra term and this is probably the
 simplest approach to study the problem on a Bethe lattice.
 
 Summarizing there are two alternative explicit approaches:
 \begin{itemize}
     \item
     We take care in an explicit way of the effects of the the birth or of the death of solutions when changing $N$ \cite{CGP5}. 
     \item We force the solutions to be stable by adding   a small perturbation \cite{RI0}.
 \end{itemize}
 Both approaches allow  explicit and correct computations and deeply clarify the physical situation. Here for reasons of
 space I cannot describe  more explicitly the  diachronic approaches and I will limit myself to the  simplest synchronic
 approach, that has the disadvantage that it cannot  be easily extended to Bethe lattice (although it could be done
 with some effort).
 
 Up to now we have seen from general arguments that, if supersymmetry is broken, there must be some zero-modes
 floating around.  Doing explicit computations for the TAP equations for the SK model it was argued
 numerically in \cite{R2} that in this case the generic stationary point has exactly one near zero modes and
 most of the stationary points come in pairs at distance $N^{-1/2}$ one from the other: they coalesce when $N
 \to \infty$ and they becomes saddles as it was argued a long time ago \cite{tap,ANDERSON} and 
 firstly seen numerically by \cite{NEMOTO}.  These results have
 been carefully verified numerically \cite{R2,CGP2}; it was finally explicitly shown that the zero mode is the
 \emph{Goldstone Fermion} of broken supersymmetry \cite{PR0} and its existence follows from the general Ward
 identities of broken supersymmetry.
 
It is possible to show that in many models there are values of the parameters where some kind of SuSy breaking
is needed \cite{R3,CLPR,AGC,CLR}.  
Let us consider the usual case, where the function $H$ itself changes in a {\sl continuous}
 way when going from $N$ to $N+1$.  For example in the case of the SK model (where
 according to the usual notation the $x$'s are called $m$) we have.
  \be
  f_{N+1}[\{m\},m(N+1)]=f_{N}[\{m\}]+\Delta f[\{m\},m(N+1)]\ ,
  \ee
 where $\{m\}$ represents the set of all $m$'s for $i=1\ldots N$, $Delta f[\{m\},m(N+1)]$ is a quantity that
 is usually of order 1, but whose derivatives with respect to $m(i)$, for $i<N+1$ vanishes when $N$ goes to
 infinity. Its explicit expression in the SK model is:
 \be
 \Delta f[\{m\},m(N+1)= L(m(N+1))-m(N+1)\sum_{k}J_{N+1,k}m(k) +\ldots
 \ee
 where $L(m)$ is a quantity of order one. The quantity $\sum_{k}J_{N+1,k}m(k)$ is also  of order 1, but each
 individual term is of order $N^{-1/2}$. Finally the dots represent other similar terms, that must be taken into
 account in the exact computation, but have characteristics similar to the terms that we have  explicitly written.
 Neglecting for simplicity  the dots (they do not change qualitatively our analysis), the equations can be written as:
 \bea
 {\partial L \over \partial m(N+1)}=\sum_{k}J_{N+1,k}m(k) \ , \\
 {\partial f_{N}[\{m\}]\over \partial m(k)}=J_{N+1,k}m(N+1)\ . \label{EQUAN}
 \eea
 The terms proportional to $J_{N+1,k}$ are of order one in the first equation, but they appears as a
 perturbation (they are of order $N^{-1/2}$) in each of the other $N$ equations.

Let us  compare what happens for a system with $N+1$ variables with a system of only
 $N$ variable. In this second case the equations are 
 \be
  {\partial f_{N}[\{m\}]\over \partial m(k)}=0 \ . \label{EQUAO}
  \ee
  Let us assume for sake of simplicity that the first equation in eq.(\ref{EQUAN}) (i.e the one  for the
  variable $m(N+1)$) has only one solution where the r.h.s is fixed.  In this case there is a one to one
  correspondence between the solutions of a system with $N+1$ variables and the solutions for a system with $N$
  variables that are obtained by perturbing the original set of equations for the $m$ variables (i.e.
  eq.(\ref{EQUAO})).  When we go  from eq.(\ref{EQUAO})to the two equations  in  (\ref{EQUAN}), solutions may
  be created or destroyed only in pairs, and this is possible only if these solutions  have
  nearly one zero mode.  The unavoidable consequence is that, if the first equation of eq.(\ref{EQUAN}) has a
  unique solution, the number of solutions may increase with $N$ only if there are small eigenvalues of the
  Hessian floating around.
  
 Therefore, if the number of solution increases exponentially with $N$, i.e. the total complexity is not zero,
 zero modes must be present some values of the parameters and they would play a role there.  The same argument
 works for external perturbations: the number of solutions of the TAP equations decreases when we increase the
 field or increase the temperature (the complexity becomes zero at large magnetic field or at large
 temperature) and therefore zero modes must be present somewhere.
 
 At the end of this analysis we find that there must always be a region where zero-modes are present: there
 our assumptions on the absence of zero modes fail and supersymmetry breaking is present.  It is difficult to
 escape the conclusion that supersymmetry is broken only for the total complexity; however this result works
 only for the total complexity: as soon as we look at the constrained complexity at $F\ne F_{T}$ \footnote{We
 denote by $F_{T}$ the value of the free energy at the maximum of the complexity, i.e. the free energy that we
 obtain if we compute the total complexity.} the argument is no more compelling.

 The following scenario is clearly possible and it is realized for example in the spherical $p$-spin models,
 as can be seen by explicit computations.  With probability 1, the solutions with free energy density less
 than $F_{T}$ have a gap in the spectrum of the Hessian, that here is totally positive: the probability of
 finding a zero mode can be computed and it is exponentially small with $N$ \cite{CGPOLD,CLR}.  Here the physical
 picture is clear: when we go from $N$ to $N+1$, the solutions at $F<F_{T}$ are stable and they only diffuse
 in free energy when $N$ is changed.  Coalescing of solutions (and the inverse phenomenon of the births of a
 pair of solutions) may happen only at $F= F_{T}$.  This complex phenomenon, which happens at $F= F_{T}$, does
 not forbid the computation of the complexity at $F\ne F_{T}$ and, by the magic hypothesis of continuity of
 the complexity as function of the free energy density, also at $F= F_{T}$.  In this way the whole computation
 of the complexity could be done at $F\ne F_{T}$ without even having to mention the creation and the
 slaughtering of the solutions that happens at $F=F_{T}$.
 
 This arcadian scenario is implemented in some models and for some time it was believed that it was the only
 possibility. However now we have realized that other scenarios are  possible and the idyllic world of spherical
 models was a rare exception.
 
 Before discussing what could happens in the various models, it is better to consider an explicit case, the SK
 model, where
 computations can be done up to the end.  This will be done in the next section. 
 
 \section{An explicit computation: the complexity of the SK model}
 
 I present here the explicit computation of the complexity in the case of TAP equations for the
 Sherrington Kirkpatrick model using the synchronic approach.  I will only consider the computation
 of the annealed complexity \cite{bm,CGMP,CLPR} and I will only mention at the end the problems related
 to the computation of the quenched complexity.
 
 I recall the formulae introduced before. The TAP free energy is given by
 \be
  f[m]=-\frac12 \sum_{i,k}m(i)J_{i,k}m(k)-\frac{\beta N}{4}(1-q^2)- T \sum_{i}S(m(i))\,  , 
  \ee
  where $q$ is a shorthand notation for $\sum_{i} m(i)^2/ N$ and $S(m)$ is the usual binary entropy.
  The corresponding equations for the stationary point are
 \be
  T\, \artanh ( m(i))- \sum_{k\ne i}  J_{i,k}m(k) - \beta m(i)(1-q) =0
 \ee
 and the matrix elements of the Hessian are given by
 \be
  \cH_{i,k}\equiv{\partial^2f\over \partial m(i)\partial m(k)}=-
  J_{i,k}+\delta_{i,k}(A(i)^{-1}+\beta(1-q))+\frac{2\beta}{N}m(i)m(k) \label{CORRECTHESSIAN}
  ,
  \ee
 If we plug the previous equations in the representation eq.(\ref{BOH}) and we neglect the term $1/N$ in the
 Hessian, (an operation legal only for the computation of the leading exponential term in $\tilde Z(w)$) we  get
\bea
 \tilde Z(w)= d\mu[X]\exp (\Omega[X] ) \nonumber \\
 \Omega[X]= \sum_{i}\lambda(i)(T \artanh ( m(i))- \sum_{k\ne i}  J_{i,k}m(k) - \beta m(i)(1-q))+ \\
 \sum_{i,k}\bar \psi(i)\psi(k) J_{i,k}+\delta_{i,k}(A(i)^{-1}+\beta(1-q))\nonumber \\
 -w\left(-\frac12 \sum_{i,k}m(i)J_{i,k}m(k)-\frac{\beta N}{4}(1-q^2)- T \sum_{i}S(m(i))\right)\ .\nonumber
 \eea
 
 We face now the problem of computing the average over the couplings $J$: quite fortunately they appear at the
 exponent in a linear way. We can thus write
 \bea
 \Omega[X]=\Omega[X]|_{J=0}+\sum_{i,k}\omega_{i,k}J_{i,k}\\
 \omega_{i,k}=\lambda(i)m(k)+ \bar \psi(i)\psi(k)- \frac{w}{2} m(i) m(k) \ .
 \eea
 The integration over the $J$ is Gaussianly trivial and we obtain 
 \be
 \tilde Z(w)=d\mu[X]\exp \left(\Omega[X]|_{J=0}+\frac12 \sum_{i,k}\left(\omega_{i,k}+\omega_{k,i}\right)^2\right)
 \ee

 In order to proceed further we must follow-the strategy of decoupling the different sites by introducing the 
 appropriate global integration parameter. At this end we can use Gaussian representations like 
 \be
 \exp\left(\frac12 \sum_{i,k}m(i)^2m(k)^2\right)\propto \int dQ \exp\left(-\frac12 Q^2+Q \sum_{i}m(i)^2\right) \ . 
 \ee
 In doing that we have to introduce many parameters $Q_{\nu,\mu}$ for $\nu,\mu=1,4$, that are respectively
 conjugate to the variables
 \be
 \sum_{i}X(i)_{\mu}X(i)_{\nu}\ ,
 \ee
 with 
 \be
 X(i)_{1}=m(i),\ \ \ X(i)_{2}=\bar \psi(i),\ \ \ X(i)_{3}=\psi(i),\ \ \ X(i)_{4}=\lambda(i) \ .
 \ee
At the end of the day we arrive at an integral over 16 variables $Q$ (there are some reductions  if we take
into accounts 
symmetries and identities, but I will not enter into these details), that has the form
\be
\int d\{Q\}\exp (N \cL(Q) )\ .
\ee
There are 8 Bosonic $Q$, that are conjugated to Bosonic variables like $m(i)^2$, $\lambda(i)m(i)$, $\bar
\psi(i)\psi(k)$ and 8 Fermionic $Q$, which are conjugated to Fermionic variables like $m(i)\psi(i)$.

The presence of Fermionic variables is crucial to make the supersymmetry manifest, but at the saddle point we 
can put the Fermionic $Q$'s equal to zero, however their presence is going to influence the corrections to the saddle point,
determining in this way the prefactor in front of the exponential \cite{PR0}.

If we eliminate the Fermions and  if we use the fact that the matrix $Q$ is symmetric and that $\psi(i)^2$ is zero, we remain
fewer   parameters $Q$. Finally we can take as parameters or the  surviving $Q$'s or their conjugated
parameters: they are
\be
 \Lambda\equiv\lan \lambda(i)^2 \ran,  \ \Delta \equiv\lan\lambda(i)m(i)\ran, \ q\equiv \lan m(i)^2 \ran,
  \ B=\lan\bar \psi(i) \psi(i)\ran -\beta (1-q) .
\ee
We can write the quantity $\cL(Q)$ by solving the the equations for the $Q$'s and expressing it as function of
$\Lambda$, $\Delta$, $q$ and $B$:
\be
\cL(\Lambda,\Delta,q,B) \ .
\ee
The value of these parameters can be obtained by imposing that the function $\cL(\Lambda,\Delta,q,B)$ is stationary with
respect to them. The final equations do have a solution with $B=0$ that turns out to be the correct one: $B=0$
implies that the diagonal elements of the inverse of the Hessian are  (in the average) the inverse of the diagonal elements
of the Hessian).

If we stick to the $B=0$ solution, one finds (numerically) that there are two others  solutions:
\begin{itemize}
    \item The supersymmetric solution where $\Lambda=\Delta=\beta(1-q)$, that depends only on one parameter.
    It can be proved that this supersymmetric solution exactly corresponds to what would be obtained using the Monasson 
    approach \cite{CGMP}.
    \item The original solution found by Bray and Moore 25 years ago, where $\Lambda\ne\Delta\ne\beta(1-q)$, that
     depends on three parameters.
 \end{itemize}
 
 It is not evident at  first sight which of the two is the correct solution, however if we compute the average
 of $\beta^2(1-m(i)^2)^2-1$ we find that it is negative on the supersymmetric solution and therefore only the
 non-supersymmetric solution is consistent.
 
 If we further compute the spectrum of the Hessian neglecting the last term in equations \ref{CORRECTHESSIAN} we
 find that it is positive definite and all eigenvalues are greater than than a quantity $G$ greater than zero 
 \cite{bm}.  Obviously this result does not
 make sense, because it implies that all the stationary points of the free energy are minima, which is in patent
 contradiction with the Morse theorem.  What   saves the day is the extra term in the Hessian
 \be
 \frac{2\beta}{N}m(i)m(k)\ .
 \ee
 An explicit computation \cite{R2} shows that while its effect is negligible as far the spectral density is concerned, it 
 can shift those few eigenvalues whose eigenvector  have a finite projection in the direction of the vector
 $m(i)$. At the and of the day one finds that there is  one eigenvalue that is shifted exactly to zero and it
 plays the role of the Goldstone Fermion of spontaneous supersymmetry breaking \cite{PR0}.
 
 The extension of these computations to the Bethe lattice is non-trivial, however, progress has been maid
  \cite{RI0,RI1,PR1}.
 
 \section{A few words on quenched complexity}
 
 The computation of the quenched complexity is much more complicated (and it is also more interesting). 
 Quenched complexity is defined as
 \be
 \Sigma_{quenched}=\overline{\Sigma_{J}}=N^{-1}\log(\#\mbox{Solutions}) \ ,
 \ee
 where the average is done over the different realizations of the system (that are labeled by $J$). Usual
 arguments tell us that the quenched complexity coincides with the typical case complexity in the large-$N$
 limit.
 
 The replica method may be used to compute the average over the control variables $J$ and everything becomes more 
 difficult.  
 Replica symmetry must be broken to perform
 the computation and it would be much more difficult to say something rigorous on the subject. 
 
 Generally speaking we have seen that the computation of the number of solutions of the TAP equations (done
 without replica symmetry breaking) corresponds, in the super symmetric case, to the evaluation of the free
 energy in the case where the replica symmetry is broken at one step.  In the quenched case, when we have to
 break replica symmetry also in the computation of the number of solutions of the TAP equations, one finds in a
 very elegant way that also here, if supersymmetry is exact, the explicit computation with $K$ levels of
 replica symmetry breaking gives the same results as the one within the Monasson approach with $K+1$ levels of
 replica symmetry breakings \cite{C4,C5}.  Unfortunately in the SK model, this result has a very limited range
 of applications because  as soon the free energy becomes greater that the the minimal free energy $F_{m}$
 the solution becomes unstable and $\beta^2(1-m(i)^2)^2$  less than 1 \cite{CLPR}.
 
 On the other hand for $F$ near to $F_{M}$ the simple supersymmetric breaking solutions seem to be OK. When $F$
 becomes smaller than $F_{c}>F_{m}$ replica symmetry must be broken and computations becomes very difficult,
 especially if also supersymmetry is broken.
 
 On the other hand it may be possible that in the case where the Monasson approach works, the computation
 could be done by a not too difficult extension of Talagrand's results.

 \section{Conclusions and open problems}
 
 Different models may behave in a different way. Let me summarize what is known at the present moment. The most
common possibilities in  the computation of the annealed complexity for a given value 
 of $F$ are the following:
 \begin{itemize}
     \item{I}: The generic stationary point is a minimum (index 0)   and  
     the modes are strictly positive: here supersymmetry is exact.
     \item{II$_{1}$}: The generic stationary point may be not a minimum.  There is a mode that has an
     eigenvalue that is nearly zero: there is a gap in the spectrum due to the other modes being greater that
     a positive constant.  The stationary points come in pairs; the two stationary points of the pair are very
     near: one has a slight positive eigenvalue and the other has a slight negative eigenvalue. In the infinite $N$
     limit the two stationary points coalesce to form a saddle point.  This is the typical situation in the
     case of broken supersymmetry \cite{AGC}.
     \item{II$_{\infty}$}: There is no gap in the spectrum: there may be one or more negative modes, but  the
     number of negative modes remains bounded when $N \to \infty$.  This typically happens at the critical
     point that separates exact supersymmetry and spontaneously broken supersymmetry
     \cite{CLR}.
     \item{III$_{\infty}$}: The number of negative modes diverges when $N \to \infty$. This usually may happen
     when $F>F_{M}$ and it is not a very-well studied case: some additional difficulties may be present as
     pointed out in \cite{potters2}.
 \end{itemize}
 For the quenched complexity there are other possibilities, due to possible breaking of replica symmetry and
 we are not going to explore this case.
 
 It evident from the Morse theorem that case I cannot be always true and in particular it cannot be true at
 $F_{M}$ where the complexity reaches its maximum.
 
 Both in cases I and II$_{1}$ the function $\Sigma(F)$ and $\tilde \Sigma(F)$ should be the same (the
 difference should be only in the prefactor), in the case II$_{\infty}$ it is possible that the two functions
 are different but in this case the value of $\tilde \Sigma(F)$ has a marginal interest.
 
 Generally speaking minima, or low index stationary points, are most likely at highly negative $F$. Therefore we expect
 that at the point $F_{m}$ where $\Sigma(F_{m})=0$ we stay in phase I. At $F_{M}$ we must stay in  phase II.
 There must be a value of $F$, $F_{c}$, that separates the two phases and it is often in the phase II$_{\infty}$.
 
 Different possibilities are realized in the region $F_{m}<F<F_{M}$;
 \begin {itemize}
 \item The whole region is in phase II$_{1}$. This is what happens in the SK model.
 \item The whole region is in phase I. This is what happens in simple $p$-spin spherical models.
 \item The region where $F_{m}<F<F_{c}$ is in phase II$_{1}$. The region where $F_{c}<F<F_{M}$ is in phase I. 
 The point $F_{c}$ is in phase II$_{\infty}$. This is what typically happens in the $p$-spin Ising models
 \cite{CLR}.
 \end{itemize}
 
 Although much progress on this field have been done in these years (especially after \cite{MP1Be,CGMP}) there are
 still many points that are not clear. I will end these lectures by mentioning a few ones:
 \begin{itemize}
     \item A careful study of the quenched case where both supersymmetry and replica symmetry are broken.
     \item A better understanding of the low-temperature phase on the Bethe lattice, especially for models
     that are of interest in combinatorial optimization.
     \item The extension of the analysis to the short range models. Here the situation would be quite
     different; real metastable states  with infinity mean life and with a  free energy density  greater than
     the equilibirum one cannot be present.
     The definition of the metastable states with a finite free energy difference with the ground state
     becomes imprecise, although this imprecision becomes smaller and smaller when the free energy density
     becomes very near
     to the equilibrium free energy. On the other hand the computation of the number of solutions of TAP like 
     equations remains well defined and it is possible that it could be easier to make progresses in this
     direction.
     \item The precise relation of the dynamics behavior  with these \emph{static} findings. More generally
     we need to start to compute barriers between different equilibrium states. A first step in this direction was
     done in \cite{CGP5}, but the results should be systematized.
     \item A systematic study of $Z(w)$ should be done, extending the results of \cite{XXX}.
\end{itemize}
As I have already  stressed many of the results presented here are not out of reach of rigorous
	  analytic computations. It would be very interesting to see them. It is quite likely that they would
	  provide a new and very useful view on the field.

\section*{Acknowledgments} 

I am very happy to have the possibility to thank all the people that have worked with me in the field: Alessia
Annibale, Andrea Cavagna, Andrea Crisanti, Silvio Franz, Irene Giardina, Luca Leuzzi, Marc M\'ezard, Andrea Montanari,
Federico Ricci-Tersenghi, Tommaso Rizzo and Elisa Trevigne.

\end{document}